\documentclass[journal]{IEEEtran}

\usepackage{graphicx}
\usepackage[caption=false,font=footnotesize]{subfig}
\usepackage{amsmath}
\usepackage{fixltx2e}%corrects a few problems
\begin{document}

\title{Statistical properties\\of chaotic binary sequences}

\author{Bogdan Cristea,~\IEEEmembership{Member,~IEEE}}

\markboth{IEEE Transactions On Information Theory}%
{Bogdan Cristea: Statistical properties of chaotic binary sequences}

\maketitle

\begin{abstract}
Mean value and cross-covariance function of chaotic binary sequences are evaluated for chaotic maps with specific properties. We also take into account the effect of fixed- and floating-point representations on statistical properties of chaotic generators. Thus, one is able to obtain possible candidates for pseudo-random binary sequences generation. Results of statistical tests applied to chaotic binary sequences are presented. The chaotic binary sequences thus obtained could be used for security improvement in IEEE 802.11 standard.
\end{abstract}


% main text
\section{Introduction}
\label{sec:intro}
Pseudo-random binary sequences have applications in several domains including information theory (informational sources), spread-spectrum systems (spreading sequences) and cryptography (stream cyphers, key generation, initialization vectors, etc.).

Chaotic generators were already studied for cryptographic applications \cite{ko01,al99,li05} due to their specific properties: sensitivity to initial conditions, random-like behavior, non-linear dynamics, etc. As already noticed in \cite{ko01}, the main difficulty when using chaotic generators in cryptography is the fact that chaotic generators take values from a continuous, infinite space, while classical encryption techniques operate on a discrete, finite space. However, since most chaotic generators are implemented using a finite precision (fixed- or floating-point), the conversion from the continuous space to the discrete space is realized implicitly \cite{li05}. Further conversion techniques allow to pass from the discrete space to a binary space thus obtaining chaotic binary sequences \cite{ko97}.

In order to obtain pseudo-random binary sequences, previously proposed approaches use a bit extraction technique so that one or several bits are extracted from each chaotic value. However, there are few studies related to the selection of chaotic maps with good statistical properties and how the bit extraction technique must be designed \cite{li05,ko97}. These are the main topics aborded in our paper. The originality of our approach resides on the fact that we take into account the effect of fixed- and floating-point representations on statistical properties of chaotic generators.

The rest of the paper is organized as follows. In section \ref{sec:mean_cov} the theoretical background used in our developments is presented. The definitions for the mean value and cross-covariance function of chaotic binary sequences are given. Subsections \ref{sec:mean} and \ref{sec:cov} give simplified expressions of mean value and cross-covariance function, respectively, for chaotic maps with specific properties. Thus one is able to obtain possible candidates for pseudo-random binary sequences generation. Results of statistical tests applied to chaotic binary sequences are presented in section \ref{sec:simu}. Finally some conclusions are given in section \ref{sec:conc}.

\section{Mean value and cross-covariance function of chaotic binary sequences}
\label{sec:mean_cov}
We consider chaotic sequences $\{x_n\}$ generated by a function (chaotic map) $\tau: [0,1]\rightarrow [0,1]$:
\begin{equation}\label{eq:chaotic_map}
 x_n = \tau(x_{n-1})
\end{equation}
where $n\in \{1,2,\ldots\}$. Each chaotic value, $x_n$, and the parameters of the chaotic map are represented using fixed- or floating-point representation.

Any chaotic map, defined on some interval $[a,b]$, $f:[a,b]\rightarrow [a,b]$, can be redefined on the interval $[0,1]$, using the tranformation: 
\begin{equation}
	\tau(x) = \frac{f(a+(b-a)x)-a}{b-a}
\end{equation}
The choice of the $[0,1]$ interval allows the simplification of our mathematical developments.

Starting from a given value $x\in [0,1]$ one can obtain the bits of the fractionary part using a sum of threshold functions \cite{ko97}:
\begin{equation}\label{eq:sum_threshold}
 b_i(x) = \sum_{r=1}^{2^i-1} (-1)^{r-1}\sigma_{\frac{r}{2^i}}(x)
\end{equation}
$i\in \{1,2,\ldots\}$. When a fixed-point representation is used for $x$, the threshold function $\sigma_t(x)$ has the expression:
\begin{equation}\label{eq:fix}
 \sigma_t(x) = \begin{cases}
                0 &, x<t\\
		1 &, x\geq t
               \end{cases}
\end{equation}
and, when a floating-point representation is used, the threshold function is:
\begin{equation}\label{eq:float}
 \sigma_t(x) = \begin{cases}
                0 &, \frac{x}{2^{e+1}}<t\\
		1 &, \frac{x}{2^{e+1}}\geq t
               \end{cases}
\end{equation}
where $e$ is the non-biased exponent of the floating-point value $x$ \cite{go91}.

The binary sequence obtained using $b_i(x)$, $i\in \{1,2,\ldots\}$ is called a chaotic binary sequence if $x$ is obtained from a chaotic generator \eqref{eq:chaotic_map}. In the following we are interested to determine under which conditions the chaotic binary sequence is pseudo-random.

For this purpose, the following statistical measures of the chaotic binary sequence are defined \cite{ko97}:
\begin{enumerate}
 \item mean value
\begin{equation}\label{eq:mean_val}
 <b_i> = \int_0^1 b_i(x) f^*(x) dx
\end{equation}
 \item cross-covariance function
\begin{equation}\label{eq:cov_fct}
\begin{split}
 <\tilde{\rho}^{(2)}(l;b_i,b_j)> &= \int_0^1 \left( b_i(x)- <b_i>\right)\\
&\left( b_j(\tau^l(x))- <b_j>\right) f^*(x) dx
\end{split}
\end{equation}
\end{enumerate}
where $f^*(x)$ is the density function of the chaotic map $\tau$ \cite[p. 8]{la94} and $\tau^l(x)$ represents the output of the chaotic map after $l$ iterations \eqref{eq:chaotic_map}: $\tau^l(x)=\tau(\tau(\cdots \tau(x)))$.

Note that, according to the above definitions, the density function is used as a probability density function. Thus, \eqref{eq:mean_val} and \eqref{eq:cov_fct} are the classical definitions of the mean and cross-covariance function of continuous random variables. In our case, the realization of the random variable is the chaotic value $x_n$, but we use a binary function $b_i(x)$ \eqref{eq:sum_threshold} in order to extract some binary information.

It is worth to emphasize that, in order to integrate in \eqref{eq:mean_val} and \eqref{eq:cov_fct}, the value $x$ must be continuous, that is represented with infinite precision. This can be accomplished when the number of bits of the fractionary part in fixed- or floating-point representations is infinite. Thus, we generalize the fixed- and floating-point representations in order to represent any number in the interval $[0,1]$. This approach allows to simplify the mathematical developments and can be seen as an ideal case of chaotic binary sequences obtained from chaotic values represented with finite precision.

Using \eqref{eq:mean_val} one is able to evaluate the mean of the binary sequence obtained from the $i$-th bit of each chaotic value generated with \eqref{eq:chaotic_map}. In order to ensure the pseudo-randomness of the chaotic binary sequence, a necessary condition is to have an (almost) equal number of zeros and ones \cite{nist01}, that is the mean value must be \eqref{eq:mean_val}:
\begin{equation}\label{eq:cond_mean}
<b_i> = 0.5
\end{equation}
for any $i$ in a given ensemble. Thus one is able to find which bits must be extracted from the fractionary part of the chaotic value in order to obtain a balanced number of zeros and ones.

Further, the cross-covariance function \eqref{eq:cov_fct} can be used to evaluate the correlations between the $i$-th and $j$-th bits belonging to different chaotic values obtained from \eqref{eq:chaotic_map}, separated by $l$ iterations. Since a non-zero cross-covariance function will imply the statistical dependence of bits, we are interested to determine when the cross-covariance function is (close to) zero \eqref{eq:cov_fct}:
\begin{equation}\label{eq:cond_cov}
<\tilde{\rho}^{(2)}(l;b_i,b_j)> = 0
\end{equation}
So, conditions \eqref{eq:cond_mean} and \eqref{eq:cond_cov} can be used to select from all available chaotic maps potential candidates for pseudo-random binary sequences generation.

In the next two subsections analytical expressions for the mean value and cross-covariance function will be deduced for chaotic maps with specific properties.

\subsection{Mean value}
\label{sec:mean}
Starting from the definition of the mean value \eqref{eq:mean_val} and using \eqref{eq:sum_threshold} the mean value can be rewritten as:
\begin{equation}\label{eq:mean_val_sum}
 <b_i> = \sum_{r=1}^{2^i-1} (-1)^{r-1} p_{\tau}(\frac{r}{2^i})
\end{equation}
where
\begin{equation}\label{eq:p_tau}
 p_{\tau}(t) = \int_0^1 \sigma_t(x) f^*(x) dx
\end{equation}

If the threshold function $\sigma_t(x)$ is defined under fixed-point representation \eqref{eq:fix}, then $p_{\tau}(t)$ \eqref{eq:p_tau} can be written as:
\begin{equation}\label{eq:p_tau_fix}
 p_{\tau}(t) = \int_t^1 f^*(x) dx
\end{equation}
and if $\sigma_t(x)$ is defined under floating-point representation, then:
\begin{equation}\label{eq:p_tau_float}
 p_{\tau}(t) = \begin{cases}
                \begin{aligned} &\int_{t2^{e_{\mathrm{min}}+1}}^{2^{e_{\mathrm{min}}}} f^*(x)dx\\& +\sum_{k=1}^{-e_{\mathrm{min}}} \int_{2^{e_{\mathrm{min}}+k-1}}^{2^{e_{\mathrm{min}}+k}} f^*(x)dx \end{aligned}
		 &, t\leq 0.5\\
		\sum_{k=1}^{-e_{\mathrm{min}}} \int_{t2^{e_{\mathrm{min}}+k}}^{2^{e_{\mathrm{min}}+k}} f^*(x)dx &, t> 0.5
               \end{cases}
\end{equation}
where $e_{\mathrm{min}}<0$ is the minimum value of the exponent $e$ (for example, in floating-point double precision $e_{\mathrm{min}}=-1022$ \cite{fp85}). In order to obtain \eqref{eq:p_tau_float} the interval $[0,1]$ is divided into sub-intervals defined by a constant exponent $e$ \cite{go91} and the definition of the threshold function $\sigma_t(x)$ \eqref{eq:float} is used.

Note that the expression of the mean value \eqref{eq:mean_val_sum} depends only of the density function, $f^*(x)$, and the representation used for the chaotic map, $\sigma_t(x)$. Thus a whole class of chaotic maps having the same density function can be characterized using \eqref{eq:mean_val_sum}.

To further simplify the expression of the mean value \eqref{eq:mean_val_sum} the density function $f^*(x)$ must be specified. We have considered two examples of density functions:
\begin{equation}\label{eq:bernoulli}f^*(x) = 1\end{equation}
and
\begin{equation}\label{eq:logistic}f^*(x) = \frac{1}{\pi \sqrt{x(1-x)}}\end{equation}
corresponding to Bernoulli map
\begin{equation}\label{eq:bernoulli_map}\tau(x)=\mathrm{mod}(px, 1)\end{equation}
where $p$ is the parameter of Bernoulli map and logistic map
\begin{equation}\label{eq:logistic_map}\tau(x)=4x(1-x)\end{equation}
respectively \cite{la94}.

Thus, it can be shown that when the density function is given by \eqref{eq:bernoulli}, under fixed-point representation, the mean value is:
\begin{equation}\label{eq:mean_fix}
 <b_i> = 0.5
\end{equation}
$i\in\{1,2,\ldots\}$ and under floating-point representation:
\begin{equation}\label{eq:mean_float}
 <b_i> = \begin{cases}
          1-2^{e_{\mathrm{min}}} &, i=1\\
	  0.5 &, i\in \{2,3,\ldots\}
         \end{cases}
\end{equation}
In order to prove \eqref{eq:mean_fix} and \eqref{eq:mean_float} see appendix.

The result given by \eqref{eq:mean_float} can be explained by the fact that, when a floating-point representation is used, the first bit $i=1$ is the hidden bit of the fractionary part (significand) \cite{go91}. Since the probability to have this bit set (normalized numbers) is much greater than the probability to have this bit unset (denormalized numbers) we obtain: $<b_1> \approx 1$. 

Following \eqref{eq:mean_fix} and \eqref{eq:mean_float} one can conclude that, when the density function is given by \eqref{eq:bernoulli}, all bits of the fractionary part (except the hidden bit) can be used to obtain balanced chaotic binary sequences.

When the density function is given by \eqref{eq:logistic}, under fixed-point representation, the mean value is:
\begin{equation}\label{eq:mean_fix2}
 <b_i> = 0.5-\frac{1}{\pi}\sum_{r=1}^{2^i-1} (-1)^{r-1} \mathrm{asin}\left( \frac{r}{2^{i-1}}-1 \right)
\end{equation}
and under floating-point representation:
\begin{equation}\label{eq:mean_float2}
 <b_i> = \begin{cases}
          0.5-\frac{1}{\pi}\mathrm{asin}\left( 2^{e_{\mathrm{min}+1}}-1 \right) &, i=1\\
	  \begin{aligned}&-\frac{1}{\pi} \sum_{r=1}^{2^{i-1}} (-1)^{r-1} \mathrm{asin}\left( r2^{e_{\mathrm{min}}-i+2}-1 \right)\\&+ \sum_{r=2^{i-1}+1}^{2^i-1} (-1)^{r-1} p_{\tau}(\frac{r}{2^i})\end{aligned} &, i>1
         \end{cases}
\end{equation}
where 
\begin{equation}
\begin{split}
 p_{\tau}(t) &= \frac{1}{\pi} \sum_{k=1}^{-e_{\mathrm{min}}} \left( \mathrm{asin}\left( 2^{e_{\mathrm{min}}+k+1}-1 \right)\right.\\
	     &- \left.\mathrm{asin}\left( t2^{e_{\mathrm{min}}+k+1}-1 \right) \right)\ \forall t>0.5
\end{split}
\end{equation}
In order to prove \eqref{eq:mean_fix2} and \eqref{eq:mean_float2} see appendix.

\begin{figure*}[ht]
	\centering
	\subfloat[][]{
	\scalebox{0.5}{\includegraphics{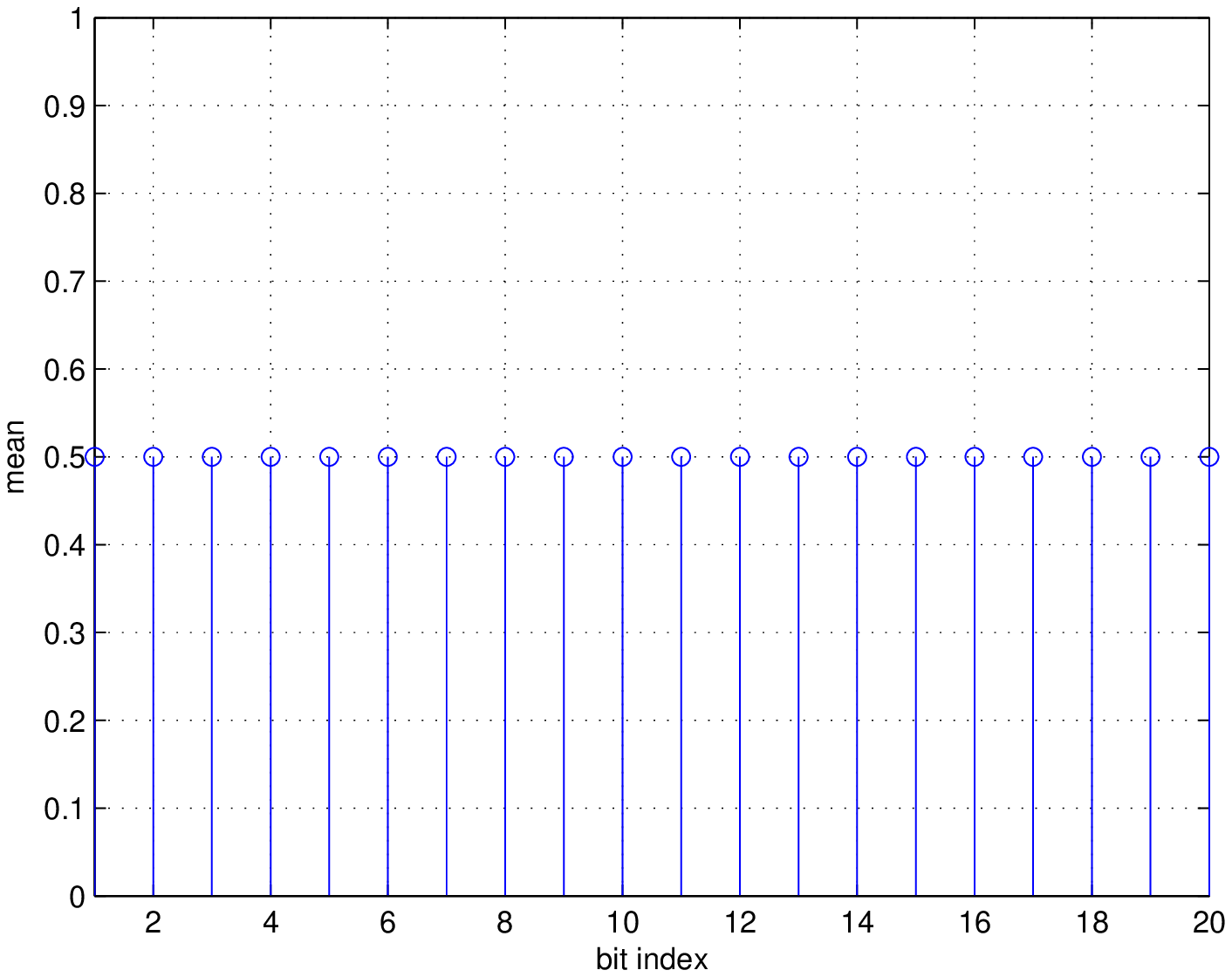}}
	\label{fig:mean_logistic_fix}}
	\subfloat[][]{
	\scalebox{0.5}{\includegraphics{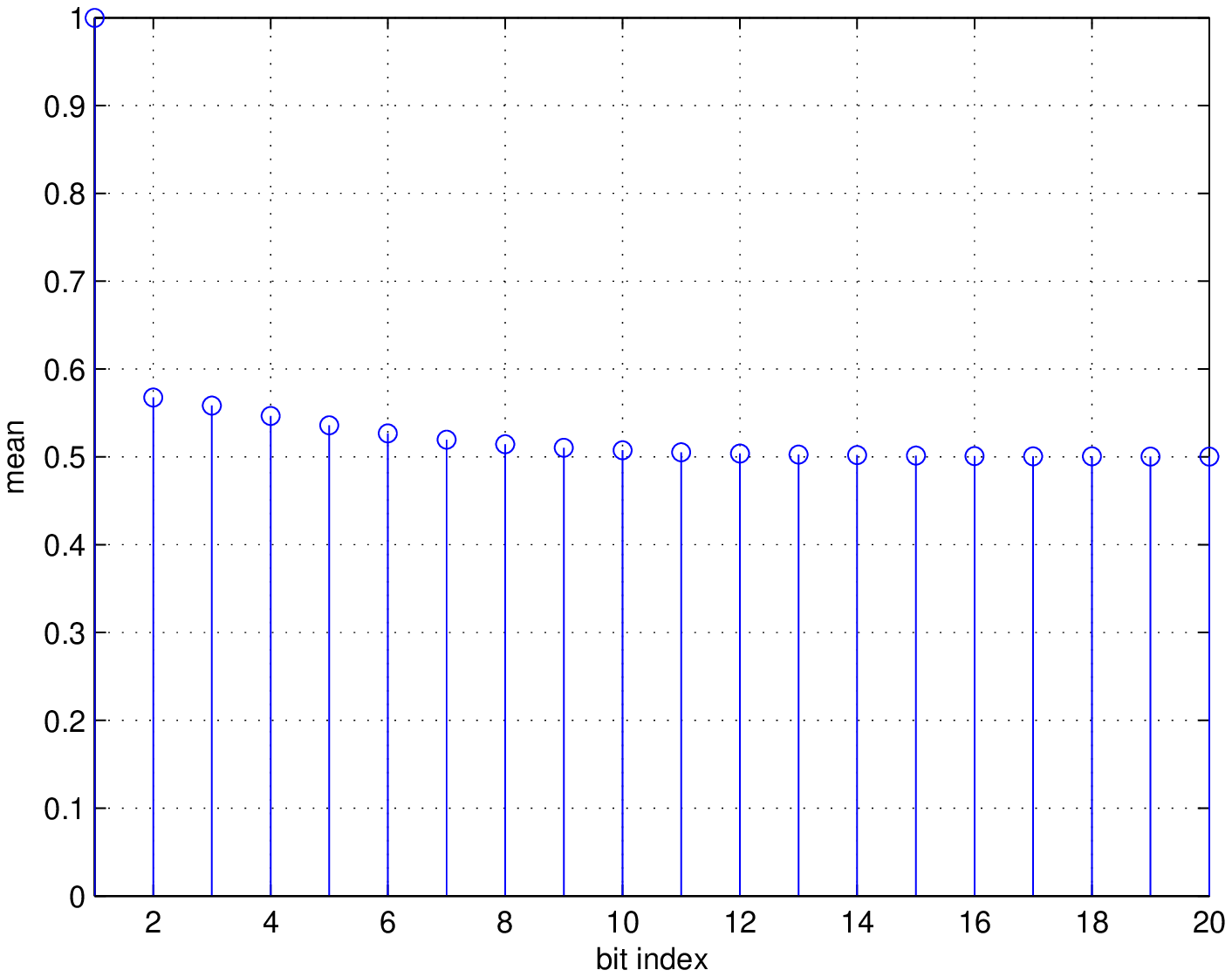}}
	\label{fig:mean_logistic_float}}
	\caption[]{Mean value corresponding to the density function given by \eqref{eq:logistic} under
	\subref{fig:mean_logistic_fix} fixed-point representation
	\subref{fig:mean_logistic_float} floating-point representation}
\end{figure*}
Numerical evaluation of relations \eqref{eq:mean_fix2} and \eqref{eq:mean_float2} shows that, when the density function is given by \eqref{eq:logistic}, under fixed-point representation, all bits of the fractionary part can be used to obtain balanced chaotic binary sequences (Fig. \ref{fig:mean_logistic_fix}). However, under floating-point representation, the mean value only converges to $0.5$ as the bit index $i$ \eqref{eq:mean_float2} increases (Fig. \ref{fig:mean_logistic_float}). So, in this case, the first bits of the fractionary part must be discarded and thus a less efficient bit extraction technique must be used in order to obtain balanced chaotic binary sequences.

So, using the mean value of the chaotic binary sequence, one is able to determine the bits of the fractionary part needed to obtain chaotic binary sequences with an equal number of zeros and ones. In order to gain further insights into the statistical properties of chaotic binary sequences, higher order statistics must be used. In the following subsection the cross-covariance function is evaluated in order to characterize the correlations at the bit level in chaotic binary sequences.

\subsection{Cross-covariance function}
\label{sec:cov}
In order to simplify the expression of the cross-covariance function \eqref{eq:cov_fct}, the chaotic map $\tau(x)$ is chosen so that \cite{ko97}:
\begin{equation}\label{eq:hyp_cov}
|g_i'(x)|f^*(g_i(x))=\frac{1}{N_{\tau}}f^*(x)\ \forall 1\leq i \leq N_{\tau}
\end{equation}
where $g_i(x)=\tau^{-1}_i(x)$, $g_i'(x)$ is the first derivative of $g_i(x)$, $\tau_i(x)$ is defined on a subinterval $[d_{i-1},d_i]\subset [0,1]$ so that $\tau_i(x)$ is invertible, $\tau_i(x)=\tau(x)\ \forall x\in [d_{i-1},d_i]$ and $N_{\tau}$ is the number of subintervals in $[0,1]$ on which $\tau(x)$ is invertible. Among the chaotic maps verifying \eqref{eq:hyp_cov} one can mention Bernoulli map, tent map, logistic map and Chebyshev polynomials \cite{ko97}.

Using \eqref{eq:hyp_cov} the cross-covariance function can be written as \cite{ko97}:
\begin{equation}\label{eq:rho_0}
\begin{split}
 <\tilde{\rho}^{(2)}(0;b_i,b_j)> &= \sum_{r=1}^{2^i-1} \sum_{s=1}^{2^j-1} (-1)^{r+s} 
\left( p_{\tau}\left( \max\left(\frac{r}{2^i},\frac{s}{2^j}\right) \right)\right.\\
&\left.-p_{\tau}\left(\frac{r}{2^i}\right)p_{\tau}\left(\frac{s}{2^j}\right) \right)
\end{split}
\end{equation}

\begin{equation}\label{eq:rho_l}
\begin{split}
 <\tilde{\rho}^{(2)}(l;b_i,b_j)> &= \frac{1}{N_{\tau}^l} \sum_{r=1}^{2^i-1} \sum_{s=1}^{2^j-1} (-1)^{r+s}\\
& \Pi_{k=1}^{l} \mathrm{sign}\left( \tau'\left( \tau^{k-1} \left(\frac{r}{2^i}\right)\right) \right)\\
& 
\left( p_{\tau}\left( \max\left(\tau^l\left(\frac{r}{2^i}\right),\frac{s}{2^j}\right) \right)\right.\\
&\left.-p_{\tau}\left(\tau^l\left(\frac{r}{2^i}\right)\right)p_{\tau}\left(\frac{s}{2^j}\right) \right)
\end{split}
\end{equation}
$\forall l>0$, where $p_{\tau}(t)$ has the expression given by \eqref{eq:p_tau}.

The cross-covariance function, \eqref{eq:rho_0} and \eqref{eq:rho_l}, depends not only of the density function, $f^*(x)$, and the representation used for the chaotic map, $\sigma_t(x)$, but also of the chaotic map itself, $\tau(x)$, and the sign of its derivative, $\mathrm{sign}\left(\tau'(x)\right)$. Note also that the cross-covariance function \eqref{eq:rho_l} is inverse proportional with the number of subintervals on which $\tau(x)$ is invertible, $N_{\tau}$. Thus, in order to have a cross-covariance function with values near to zero, the chaotic map must be chosen so that $N_{\tau}$ is large enough. For example, the Bernoulli map \eqref{eq:bernoulli_map} with $p=1.99$ has $N_{\tau}=2$, but when $p=3.99$ we have $N_{\tau}=4$.

\begin{figure*}[ht]
	\centering
	\subfloat[][]{
	\scalebox{0.5}{\includegraphics{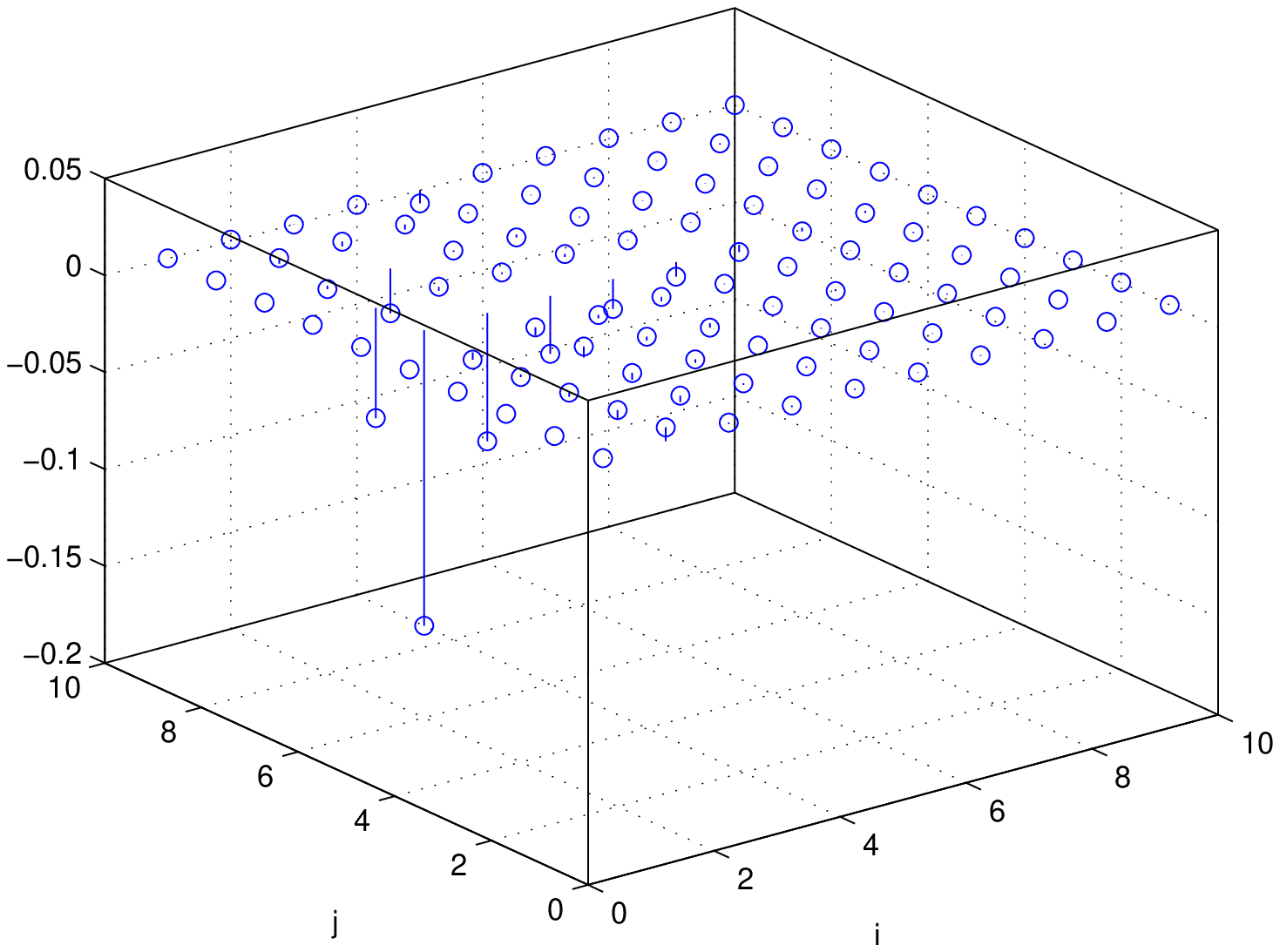}}
	\label{fig:cov_bernoulli2_float}}
	\subfloat[][]{
	\scalebox{0.5}{\includegraphics{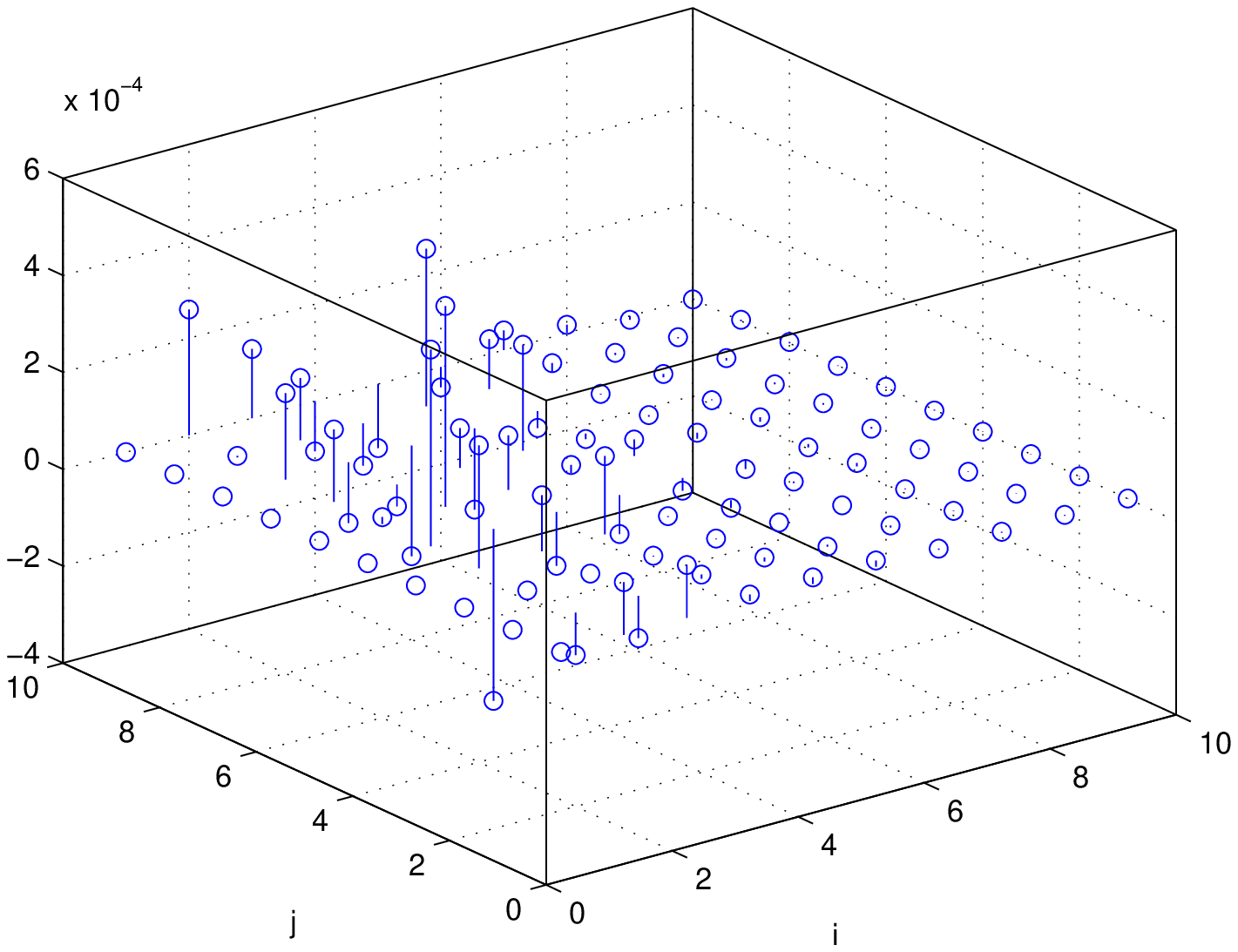}}
	\label{fig:cov_bernoulli4_float}}
	\caption[]{Cross-covariance function for $l=5$ corresponding to the Bernoulli map \eqref{eq:bernoulli_map} under floating-point representation with
	\subref{fig:cov_bernoulli2_float} $p=1.99$
	\subref{fig:cov_bernoulli4_float} $p=3.99$}
\end{figure*}

Numerical evaluation of \eqref{eq:rho_l} for Bernoulli map under floating-point representation is given in Fig \ref{fig:cov_bernoulli2_float} and \ref{fig:cov_bernoulli4_float}. It can be seen that, when the number of iterations $l$ is fixed, the correlations between the bits of different chaotic values can be reduced by choosing the parameter of the Bernoulli map so that $N_{\tau}$ is large enough.

However, the expression of the cross-covariance function is too complex to provide insights into statistical properties of chaotic maps without full specification of the chaotic map. Since our goal is to find pseudo-random generators, we have found more useful to direclty apply statistical tests \cite{nist01} on chaotic binary sequences. The results of the application of statistical tests are given in the following section.

\section{Simulation results}
\label{sec:simu}
In our simulations we use the chaotic maps defined by \eqref{eq:bernoulli_map} and \eqref{eq:logistic_map}, the tent map
\begin{equation}
 \tau(x) = p(1-|1-2x|)
\end{equation}
where $p=0.99$ and the Chebyshev polynomial of third order
\begin{equation}
 \tau_p(x) = (2x-1)(2\tau_{p-1}(x)-1)+1-\tau_{p-2}(x)
\end{equation}
where $\tau_1(x)=x$, $\tau_0(x)=1$ and $p=3$.

Since under a finite representation (fixed- or floating-point) the chaotic sequences are periodic, we use in our simulations initial conditions generating chaotic sequences with the longest period \cite{cr07b}. The parameters used in our simulations are synthesized in table \ref{tab:simu}. The fixed-point representation on $32$ bits is defined by one bit for the sign, $3$ bits for the integer part and $28$ bits for the fractionary part. The floating-point representation on $32$ bits is the IEEE754 floating-point single-precision representation \cite{fp85}. We use also the floating-point double-precision representation with the same initial conditions as for single-precision (this choice is due to the fact that in double-precision finding the longest period takes a much longer time than for single-precision). From each chaotic value, represented with finite precision, the bits of the fractionary part (except the hidden bit when floating-point representation is used) are extracted and thus chaotic binary sequences are obtained.

\begin{table}
\centering
\begin{tabular}{c|c|c|c}
 chaotic map & representation & initial condition & period len. \\
\hline \hline
Bernoulli map &	fixed ($32$ bits)	   & $0.58268645778298\ldots$ & $3050$\\
($p=3.99$)	      & float ($32$ bits) & $0.24101102352142\ldots$ & $3827$\\
\hline
tent map &	fixed ($32$ bits)	   & $0.0453341640532\ldots$ & $583$\\
	      & float ($32$ bits) & $0.4008057713508\ldots$ & $4311$\\
\hline
logistic map &	fixed ($32$ bits)	   & $0.0009547546505\ldots$ & $13404$\\
	      & float ($32$ bits) & $0.99998587369918\ldots$ & $930$\\
\hline
Chebyshev pol. &	fixed ($32$ bits)	   & $0.99994068592786\ldots$ & $10014$\\
of third order        & float ($32$ bits) & $0.05464851856231\ldots$ & $1821$
\end{tabular}
\caption{Simulation parameters}
\label{tab:simu}
\end{table}

Using chaotic sequences represented with floating-point double precision, we have numerically computed the probability density function for each chaotic map. Thus, we have found that Bernoulli map and tent map belong to the same class of chaotic maps having the density function given by \eqref{eq:bernoulli}. Logistic map and Chebyshev polynomial of third order have a density function close to \eqref{eq:logistic}. So, the results obtained for the mean value using \eqref{eq:bernoulli} and \eqref{eq:logistic} are appliable for tent map and Chebyshev polynomial of third order, respectively.

\begin{figure*}[ht]
	\centering
	\scalebox{0.8}{\includegraphics{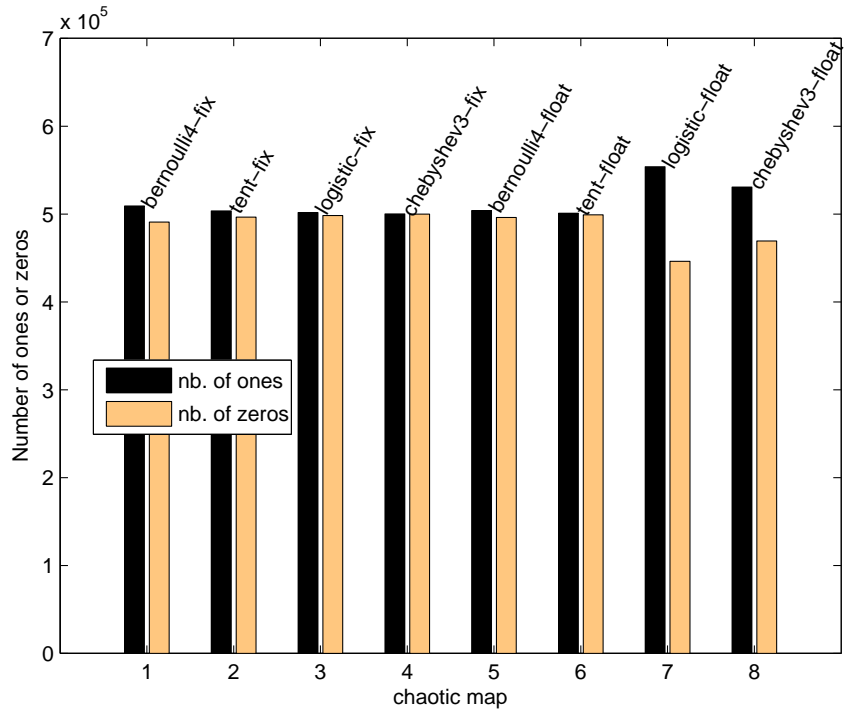}}
	\label{fig:frequency}
	\caption{Number of zeros and ones of chaotic binary sequences obtained from several chaotic maps}
\end{figure*}
In Fig. \ref{fig:frequency} we have computed the number of zeros and ones of chaotic binary sequences with $10^6$ bits. Thus, we have found that logistic map and Chebyshev polynomial of third order under single-precision floating-point representation generate chaotic binary sequences with a greater number of ones than zeros. This result is in agreement with the theoretical result obtained for the mean value \eqref{eq:mean_float2} (Fig. \ref{fig:mean_logistic_float}).Using the same representation, Bernoulli map and tent map have an almost equal number of zeros and ones \eqref{eq:mean_float}. When the fixed-point representation is used, all considered chaotic maps generate balanced chaotic binary sequences. The differences with respect to theoretical results, \eqref{eq:mean_fix}, \eqref{eq:mean_float} and \eqref{eq:mean_fix2}, are due to the fact that, in simulation, we use a finite precision, while in theory we have supposed that the number of bits of the fractionary part is infinite. However, if the number of bits of the fractionary part is large enough, we find a good agreement between theory and simulation.

\begin{figure*}[ht]
	\centering
	\subfloat[][]{
	\scalebox{0.5}{\includegraphics{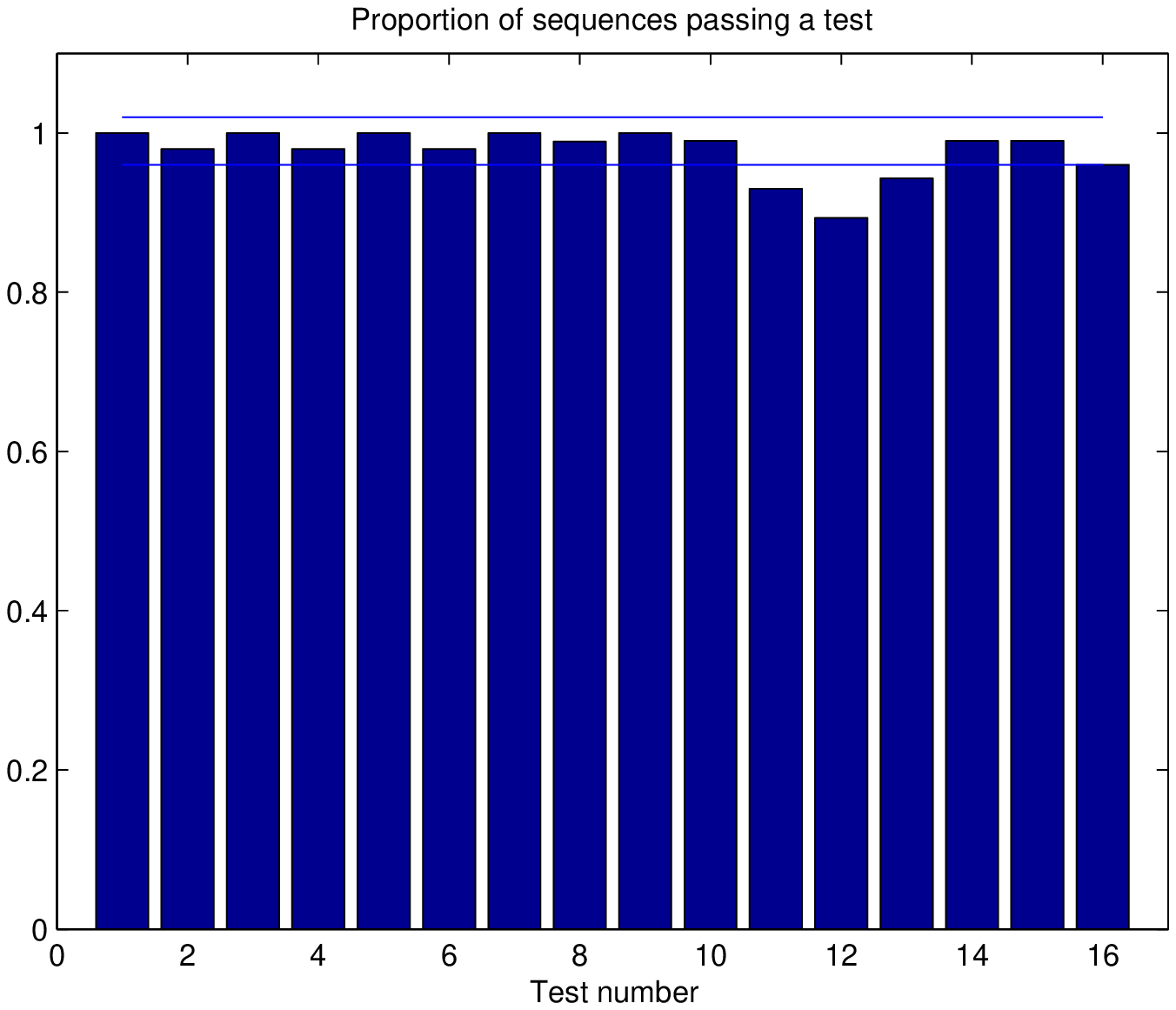}}
	\label{fig:nist_r4adic_double}}
	\subfloat[][]{
	\scalebox{0.5}{\includegraphics{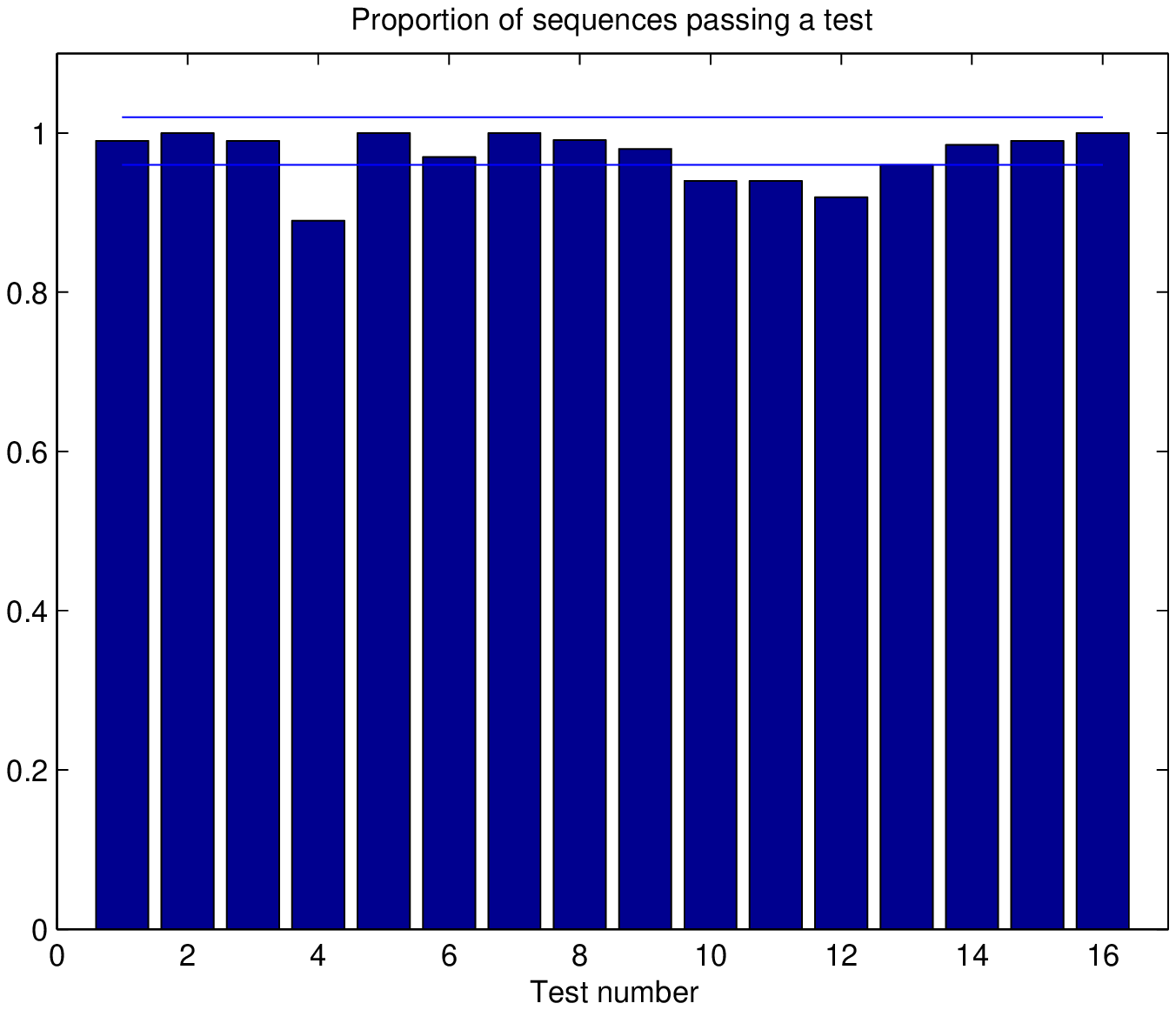}}
	\label{fig:nist_tent_double}}
	\caption[]{Statistical tests results under floating-point double precision representation for
	\subref{fig:nist_r4adic_double} Bernoulli map, $p=3.99$
	\subref{fig:nist_tent_double} tent map}
\end{figure*}

We have also applied all $16$ statistical tests described in \cite{nist01} on chaotic binary sequences generated with the above described chaotic maps. We have used $100$ binary sequences of $10^6$ bits each. The best results were obtained with Bernoulli map (Fig. \ref{fig:nist_r4adic_double}) and tent map (Fig. \ref{fig:nist_tent_double}) under floating-point double precision representation.

Thus, the theoretical results given in subsections \ref{sec:mean} and \ref{sec:cov} have allowed to select chaotic maps suitable for pseudo-random sequences generation. Further, one can select the bits to be extracted from each chaotic value for pseudo-random sequences generation. Bernoulli map and tent map were selected since chaotic maps with density function given by \eqref{eq:bernoulli} generate balanced chaotic binary sequences using all bits of the fractionary part. Note that the parameter of Bernoulli map $p=3.99$ was chosen based on the fact that the cross-covariance function has smaller values when $N_{\tau}=4$. 

By simulation, we have noticed that the period length of chaotic sequences has an impact on the results of statistical tests. The longer the period is, the better are the results of statistical tests. This observation could provide another explanation why the best results were obtained using double-precision floating-point representation (Fig. \ref{fig:nist_r4adic_double} and \ref{fig:nist_tent_double}).

Instead of selecting initial conditions generating chaotic sequences with the longest period, one can use arbitrary initial conditions and the perturbations technique \cite{zh97b}. Our simulations have shown that, with respect to our method, the perturbations technique brings little improvements of the results of statistical tests. However, the perturbations technique represents a practical approach for the implementation of chaotic pseudo-random generators e.g. in the Wired Equivalent Privacy (WEP) protocol of IEEE 802.11 standard \cite{ga05}.

The WEP protocol is particulary interesting for the application of chaotic generators since the initialization vector (initial condition) is often changed. Thus, short length periods or fixed points, specific for chaotic maps represented with finite precision, can be avoided. We expect that security improvements in IEEE 802.11 standard could be achieved by replacing the RC4 pseudo-random generator with chaotic binary generators.

\section{Conclusion}
\label{sec:conc}
The mean value and cross-covariance function of chaotic binary sequences were analytically computed. The considered chaotic sequences were represented using fixed- and floating-point representations. We have shown that chaotic binary sequences can produce binary sequences with an equal number of zeros and ones if the density function is $f^*(x)=1$. Further, these chaotic binary sequences were obtained using all the bits of the fractionary part of the chaotic value. This bit extraction technique is highly efficient with respect to previously proposed methods. Improvements of correlation properties of chaotic binary sequences are possible by increasing the number of subintervals, $N_{\tau}$, on which the chaotic map is invertible. Using these results, possible candidates (chaotic maps) for pseudo-random binary sequences generation can be selected. We have also evaluated, by means of statistical tests, chaotic binary sequences thus obtained.

Further research needs to be conducted for chaotic maps under fixed-point representation. We have considered a relatively short fixed-point representation on $32$ bits, but we expect that increasing the precision in this case could provide also good pseudo-random binary sequences. Another interesting research direction is to find out how much precision is needed in order to obtain chaotic binary sequences with good statistical properties. Also, it should be interesting to develop statistical measures by taking into account the finite representation of chaotic values, that is by replacing in the definitions of the mean value and cross-covariance function the integral by a sum. 

A possible application of chaotic binary sequences could be the WEP protocol of IEEE 802.11 standard by replacing the RC4 pseudo-random generator with chaotic binary generators. However, further cryptanalysis research should be done in order to prove the validity of this approach.

\appendix

\begin{description}
 \item[i.] \textbf{Mean value for $f^*(x)=1$}
\end{description}

%\section{Mean value for $f^*(x)=1$}

Fixed-point representation

Knowing that \eqref{eq:p_tau_fix}: 
\begin{equation}
	p_{\tau}(t)=1-t
\end{equation}
we get \eqref{eq:mean_val_sum}:
\begin{equation}\label{eq:mean_fix1}
 <b_i> = \sum_{r=1}^{2^i-1} (-1)^{r-1} (1-\frac{r}{2^i})
\end{equation}
With
\begin{equation}
 \sum_{r=1}^{n}(-1)^{r-1}r=(-1)^{n-1}\left \lceil \frac{n}{2} \right \rceil
\end{equation}
and \eqref{eq:mean_fix1} we get \eqref{eq:mean_fix}.

Floating-point representation

Knowing that \eqref{eq:p_tau_float}:
\begin{equation}
 p_{\tau}(t) = \begin{cases}
                2^{e_{\mathrm{min}}}(1-2t)+1-2^{e_{\mathrm{min}}} &, t\leq 0.5\\
		2(1-t)(1-2^{e_{\mathrm{min}}}) &, t>0.5
               \end{cases}
\end{equation}
we get \eqref{eq:mean_val_sum}:

if $i=1$
\begin{equation}
 <b_i> = p_{\tau}(\frac{1}{2}) = 1-2^{e_{\mathrm{min}}}
\end{equation}

if $i>1$
\begin{align}
 <b_i> &= \sum_{r=1}^{2^{i-1}}(-1)^{r-1} p_{\tau}(\frac{r}{2^i})\\
&\quad+\sum_{r=2^{i-1}+1}^{2^i-1}(-1)^{r-1} p_{\tau}(\frac{r}{2^i})\\
&= \frac{1}{2}
\end{align}
So we have completely proved \eqref{eq:mean_float}.

\begin{description}
 \item[ii.] \textbf{Mean value for $f^*(x)=\frac{1}{\pi \sqrt{x(1-x)}}$}
\end{description}
The following result is used:
\begin{equation}\label{eq:asin}
 \int \frac{1}{\pi \sqrt{x(1-x)}} dx = \frac{1}{\pi} \mathrm{asin}\left( 2x-1 \right)
\end{equation}

Fixed-point representation

With \eqref{eq:asin} and \eqref{eq:p_tau_fix} we get:
\begin{equation}
 p_{\tau}(t) = 0.5-\frac{1}{\pi} \mathrm{asin}\left( 2t-1 \right)
\end{equation}
So \eqref{eq:mean_val_sum}:
\begin{equation}
 <b_i> = 0.5-\frac{1}{\pi}\sum_{r=1}^{2^i-1} (-1)^{r-1} \mathrm{asin}\left( \frac{r}{2^{i-1}}-1 \right)
\end{equation}

Floating-point representation

with \eqref{eq:asin} and \eqref{eq:p_tau_float} we get:
\begin{equation}
 p_{\tau}(t) = \begin{cases}
                0.5-\frac{1}{\pi} \mathrm{asin}\left( t2^{e_{\mathrm{min}}+2}-1 \right) &, t\leq 0.5\\
		\begin{aligned}&\frac{1}{\pi}\sum_{k=1}^{-e_{\mathrm{min}}} \left( \mathrm{asin}\left( 2^{e_{\mathrm{min}}+k+1}-1 \right)\right.\\&\left. - \mathrm{asin}\left( t2^{e_{\mathrm{min}}+k+1}-1 \right) \right)\end{aligned} &, t>0.5
               \end{cases}
\end{equation}
So \eqref{eq:mean_val_sum}:

If $i=1$
\begin{equation}
 <b_i> = p_{\tau}(\frac{1}{2}) = 0.5-\frac{1}{\pi}\mathrm{asin}\left( 2^{e_{\mathrm{min}+1}}-1 \right)
\end{equation}

If $i>1$
\begin{align}
 <b_i> &= \sum_{r=1}^{2^{i-1}}(-1)^{r-1} p_{\tau}(\frac{r}{2^i})+\sum_{r=2^{i-1}+1}^{2^i-1}(-1)^{r-1} p_{\tau}(\frac{r}{2^i})\\
&= -\frac{1}{\pi} \sum_{r=1}^{2^{i-1}} (-1)^{r-1} \mathrm{asin}\left( r2^{e_{\mathrm{min}}-i+2}-1 \right)\\
&\quad+ \sum_{r=2^{i-1}+1}^{2^i-1} (-1)^{r-1} p_{\tau}(\frac{r}{2^i})
\end{align}
So we have completely proved \eqref{eq:mean_float2}.


\newpage
\begin{thebibliography}{10}

\bibitem{ko01}
L.~Kocarev, ``Chaos-based cryptography: a brief overview,'' {\em Circuits and
  Systems Magazine, IEEE}, vol.~1, no.~3, pp.~6--21, 2001.

\bibitem{al99}
G.~Alvarez, P.~Montoya, G.~Pastor, and M.~Romera, ``Chaotic cryptosystems,'' in
  {\em Security Technology, 1999. Proceedings. IEEE 33rd Annual 1999
  International Carnahan Conference on}, pp.~332--338, 1999.

\bibitem{li05}
S.~Li, G.~Chen, and X.~Mou, ``On the dynamical degradation of digital piecewise
  linear chaotic maps,'' {\em International Journal of Bifurcation and Chaos},
  vol.~15, no.~10, pp.~3119--3151, 2005.

\bibitem{ko97}
T.~Kohda and A.~Tsuneda, ``Statistics of chaotic binary sequences,'' {\em
  {IEEE} Trans. Inf. Theory}, vol.~43, pp.~104--112, Jan. 1997.

\bibitem{go91}
D.~Goldberg, ``What every computer scientist should know about floating-point
  arithmetic.'' Computing surveys, Mar. 1991.

\bibitem{la94}
A.~Lasota and M.~C. Mackey, {\em Chaos, fractals and noise}.
\newblock New-York: Springer-Verlag, 1994.

\bibitem{nist01}
A.~Rukhin, J.~Soto, J.~Nechvatal, M.~Smid, E.~Barker, S.~Leigh, M.~Levenson,
  M.~Vangel, D.~Banks, A.~Heckert, J.~Dray, and S.~Vo, ``A statistical test
  suite for pseudoradom number generators used in cryptographic applications,''
  tech. rep., National Institute of Standards and Technology, 2001.

\bibitem{fp85}
J.~E. May, J.~E. Riganati, and S.~E. Sherr, ``{IEEE} standard for binary
  floating-point arithmetic.'' American National Standards Institute, July
  1985.

\bibitem{cr07b}
B.~Cristea, P.~Charge, D.~Fournier-Prunaret, F.~Peyrard, and J.-J. Mercier,
  ``Behavior of chaotic sequences under a finite representation and its
  cryptographic applications,'' {\em IEEE Workshop on nonlinear maps and
  applications (NOMA'07)}, December 2007.
\newblock {T}oulouse, {F}rance.

\bibitem{zh97b}
H.~Zhou and X.-T. Ling, ``Realizing finite precision chaotic systems via
  perturbation of m-sequences.'' Acta Electronica Sinica, 1997.

\bibitem{ga05}
M.~Gast, {\em 802.11 Wireless networks. The definitive guide}.
\newblock O'Reilly, 2005.

\end{thebibliography}
\end{document}